\documentclass[10pt,aps,pra,twocolumn,floatfix,nofootinbib,superscriptaddress,longbibliography]{revtex4-2}
\usepackage[utf8]{inputenc}

\usepackage[dvipsnames]{xcolor}
\definecolor{alecolor}{RGB}{198,113,190}
\usepackage{bm,graphicx,mathrsfs,amsmath,amssymb,mathtools,makecell,bbm,amsthm,dsfont,color,times,txfonts,nicefrac,framed,enumitem,tikz,physics,wrapfig,amsfonts,lipsum,nicematrix,xargs, multirow}
\definecolor{equationcolor}{RGB}{222,94,100}
\definecolor{changescolor}{rgb}{0, 0, 0.7}

\usepackage[colorlinks=true,linkcolor=alecolor,citecolor=equationcolor]{hyperref}

\hypersetup{urlcolor=equationcolor}

\definecolor{equationcolor}{RGB}{154,135,198}
\usetikzlibrary{arrows.meta}
\usetikzlibrary{positioning}
\usepackage{float}
\usepackage{placeins}
\usetikzlibrary{calc} 
\usepackage{ragged2e}
\usepackage{comment}

\begin{document}

\title{Harnessing dimensional leakage for continuous-variable randomness certification}

\author{Moisés Alves}
\thanks{These authors contributed equally to this work.}
\affiliation{Center for Quantum-Enabled Computing, Center for Theoretical Physics, Polish Academy of Sciences, al. Lotników 32/46, 02-668 Warsaw, Poland}
\affiliation{International Institute of Physics, Federal University of Rio Grande do Norte, 59078-970, Natal, Brazil}
\affiliation{QuIIN - Quantum Industrial Innovation, EMBRAPII CIMATEC Competence Center in Quantum Technologies, SENAI CIMATEC, Av. Orlando Gomes 1845, 41650-010, Salvador, BA, Brazil}
\author{Tailan S. Sarubi}
\thanks{These authors contributed equally to this work.}
\affiliation{Physics Department, Federal University of Rio Grande do Norte, Natal, 59072-970, Rio Grande do Norte, Brazil}
\affiliation{International Institute of Physics, Federal University of Rio Grande do Norte, 59078-970, Natal, Brazil}
\author{Vitor L. Sena}
\affiliation{QuIIN - Quantum Industrial Innovation, EMBRAPII CIMATEC Competence Center in Quantum Technologies, SENAI CIMATEC, Av. Orlando Gomes 1845, 41650-010, Salvador, BA, Brazil}
\author{Alexandre B. Tacla}
\affiliation{QuIIN - Quantum Industrial Innovation, EMBRAPII CIMATEC Competence Center in Quantum Technologies, SENAI CIMATEC, Av. Orlando Gomes 1845, 41650-010, Salvador, BA, Brazil}
\author{Rafael Chaves}
\affiliation{International Institute of Physics, Federal University of Rio Grande do Norte, 59078-970, Natal, Brazil}


\begin{abstract}
Semi-device-independent quantum protocols often rely on strict Hilbert-space dimension assumptions that are difficult to satisfy in realistic implementations, particularly with infinite-dimensional optical fields. The almost-qudit framework addresses this issue by quantifying the leakage of quantum states beyond a target subspace. Although such leakage can severely compromise randomness certification for fixed preparation and measurement inputs, we show that averaging over the full input ensemble can mitigate this vulnerability and extend the range of tolerable leakage. This enhanced robustness enables nonzero randomness certification using low-squeezing single-mode Gaussian states and homodyne detection, thus offering an experimentally accessible continuous-variable implementation.
\end{abstract}

\maketitle

\section{Introduction}
\label{sec:introducao}

Prepare-and-measure (PM) scenarios~\cite{brask2026quantum,junior2026quantum} provide a versatile framework for quantum information processing when the preparation and measurement devices are not fully characterized. By assuming that the communicated quantum system satisfies physical constraints~\cite{van2017semi,Rusca2019,Avesani2021,Roch2025,Brask2017,Shi2019,Tavakoli2021,chaves2015device,sarubi2026detection,Tavakoli2020Info,Tavakoli2022Info,cobucci2026operationally,Pauwels2025information}, most notably a bounded Hilbert-space dimension~\cite{gallego2010,sarubi2026prepare,NavascuesVertesi2015,Bernal2024absolute}, semi-device-independent protocols can certify quantum properties without requiring a detailed model of the underlying devices. These include quantum randomness generation~\cite{pawlowski2011,Li2012,alves2026semi,sarubi2026randomness,Passaro2015,Mironowicz2016,Roch2025EAT}, and the certification and characterization of properties of quantum states and measurements~\cite{tavakoli2018self,tavakoli2020self,miklin2020semi,zamora2025semi,Navascues2023self}. Such protocols are attractive for practical quantum technologies, where a full device-independent analysis may be experimentally demanding~\cite{scarani2012device,acin2016}. Their reliability, however, frequently rests on the assumption that the physical systems used in the experiment satisfy the prescribed dimension constraint~\cite{Pauwels2022}.

In realistic implementations, this assumption is rarely exact. Optical fields, for instance, occupy infinite-dimensional Hilbert spaces~\cite{Weedbrook2012,Brask2022}, and even when information is encoded in a nominal qubit or qudit, such as in the polarization or orbital angular momentum of photons, multiphoton contributions, higher-order Fock-state occupations, and other unwanted degrees of freedom can extend beyond the intended subspace. To address this issue, the concept of almost qudits has been introduced~\cite{Pauwels2022}, in which a quantum state is predominantly supported on a target $d$-dimensional subspace but may contain a small higher-dimensional component. Specifically, a state $\rho_x$ associated with preparation $x$ is an almost qudit if
\begin{equation}\label{EQ1}
\operatorname{Tr}(\rho_x\Pi_d)\geq 1-\varepsilon_x,
\end{equation}
where $\Pi_d$ projects onto the target $d$-dimensional subspace and $\varepsilon_x$ quantifies the dimensional leakage. This operational description accommodates physical systems while retaining the central premise of dimension-based semi-device-independent protocols.

The introduction of almost qudits also revealed a severe limitation of these protocols, since even a small dimension leakage can substantially compromise certification procedures~\cite{Pauwels2022,Pauwels2024erratum}, including quantum randomness generation and measurement certification. In randomness generation, the problem is particularly pronounced when one seeks to certify the output associated with a specific preparation and measurement setting. An adversary with classical side information can exploit the additional degrees of freedom to reproduce the observed statistics while making the selected output more predictable than an ideal-qudit analysis would suggest. Consequently, the certified randomness can deteriorate rapidly as the dimension leakage increases, even when the system remains predominantly confined to the intended subspace. This sensitivity raises an important question: to what extent does the fragility of existing protocols reflect a fundamental limitation imposed by dimensional leakage, and to what extent does it depend on how the protocol's performance is evaluated?

In this work, we show that averaging the guessing probability over preparation and measurement inputs can substantially mitigate the effects of dimensional leakage. Unlike fixed-input certification, we exploit the full input ensemble, yielding more robust randomness guarantees and extending the range of leakage parameters compatible with nonzero certified randomness. We further demonstrate that this robustness enables practical continuous-variable (CV) implementations using Gaussian states with reasonably low squeezing and homodyne measurements, which have also been used in other quantum randomness generation schemes~\cite{gehring2021,Ishihara2026}. Rather than requiring strictly finite-dimensional states, our framework accommodates their higher-dimensional components, turning the tolerance of dimensional leakage from a potential limitation into a practical advantage for experimentally accessible semi-device-independent randomness generation.

\begin{figure*}[t]
    \centering
    \includegraphics{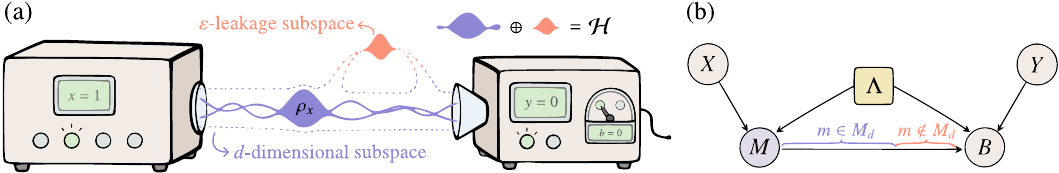}
    \caption{
    \textbf{Almost-$d$-dimensional prepare-and-measure scenario.}
    (a) Quantum realization. The preparation device receives
    an input $x$ and emits a state in a Hilbert space
    $\mathcal{H}$. The measurement device receives the
    system and an input $y$, and produces an outcome $b$.
    For each preparation, the population outside the
    designated $d$-dimensional subspace, averaged over
    shared classical randomness, is at most $\varepsilon$.
    The dependence of the states, measurements, and
    subspace on the shared variable is omitted in the
    diagram.
    (b) Classical realization. The input $X$ is encoded
    into a message $M$, which, together with the
    measurement input $Y$, determines the output $B$.
    The shared variable $\Lambda$ may correlate the
    devices. For each preparation, the probability that
    the message lies outside the designated alphabet
    of at most $d$ symbols, averaged over $\Lambda$,
    is at most $\varepsilon$.
    }
    \label{fig:PAMdiagram}
\end{figure*}

The remainder of the paper is organized as follows. Sec.~\ref{sec:pam} introduces the almost-qudit PM scenario. Sec.~\ref{sec:average_randomness} presents the randomness certification method and compares the average and fixed-input estimators. Sec.~\ref{sec:cv_implementations} explores CV implementations using Gaussian and non-Gaussian states with homodyne and displaced photodetection. Sec.~\ref{sec:conclusao} concludes the paper. The Appendices~\ref{app:main_scenarios}, \ref{app:classical}, and \ref{app:gaussian_fock} provide results for additional witnesses and technical derivations.

\section{Almost-\texorpdfstring{$d$}{d}-dimensional prepare-and-measure scenario}
\label{sec:pam}

The prepare-and-measure scenario consists of a preparation device that receives an input $x\in\{0,\dots,|X|-1\}$ and sends a physical system to a measurement device. The latter
receives an input $y\in\{0,\dots,|Y|-1\}$ and returns an outcome $b\in\{0,\dots,|B|-1\}$. The resulting statistics define a behavior $\{p(b|x,y)\}_{b,x,y}$. In a classical
description, the communicated system carries a message $m\in\mathcal{M}$, implying that
\begin{equation}
    p(b|x,y)
    =
    \sum_{\lambda,m}q(\lambda)
    p(m|x,\lambda)p(b|m,y,\lambda),
    \label{eq:classical_behavior}
\end{equation}
where $\lambda$ represents shared classical randomness, independent of the inputs. In a quantum description, the message is replaced by a state $\rho_x^\lambda$ and the measurement by a POVM $\{M_{b|y}^\lambda\}_b$, so that
\begin{equation}
    p(b|x,y)
    =
    \sum_{\lambda}q(\lambda)
    \operatorname{Tr}
    \left(\rho_x^\lambda M_{b|y}^\lambda\right).
    \label{eq:quantum_shared_randomness}
\end{equation}

A separation between classical and quantum correlations
requires a restriction on the communicated system.
The standard dimension assumption imposes
$|\mathcal{M}|\leq d$ classically and
$\dim(\mathcal{H})\leq d$ quantum
mechanically~\cite{gallego2010,pawlowski2011,Li2012}.
Alternative assumptions include bounds on the energy
or mean photon number~\cite{van2017semi,Rusca2019,Avesani2021},
constraints on photon-number statistics~\cite{Roch2025},
on overlaps between preparations~\cite{Brask2017,Shi2019},
on fidelities with reference states~\cite{Tavakoli2021},
or on the information carried by the communicated
system~\cite{chaves2015device,Tavakoli2020Info,Tavakoli2022Info,mishra2026}.

The almost-qudit framework relaxes the dimension assumption by bounding the population outside a $d$-dimensional subspace, while leaving the ambient Hilbert space unrestricted~\cite{Pauwels2022}, as expressed by Eq.~\eqref{EQ1} and illustrated schematically in Fig.~\ref{fig:PAMdiagram}(a). Here we assume the same leakage for all preparations, $\varepsilon_x=\varepsilon$, for simplicity. The $d$-dimensional subspace is defined by a rank-$d$ projector $\Pi_d=\sum_{j=0}^{d-1}|j\rangle\langle j|$, for some orthonormal basis $\{|j\rangle\}_{j=0}^{d-1}$, common to all preparations. For $\varepsilon=0$, the usual dimension-bounded scenario is recovered, while for $\varepsilon>0$ the preparations may have some population outside the intended $d$-dimensional subspace.

In the presence of shared randomness, Eq.~\eqref{EQ1} can be extended in two natural ways. One possibility is to require the leakage bound for every realization $\lambda$,
\begin{equation}
    \operatorname{Tr}
    \left(\rho_x^\lambda\Pi_d^\lambda\right)
    \geq 1-\varepsilon
    \qquad \forall x,\lambda,
    \label{eq:branchwise_leakage}
\end{equation}
where $\Pi_d^\lambda$ is a rank-$d$ projector common to all preparations associated with a given $\lambda$. We use, however, a less restrictive alternative, imposing the same condition only on average over the shared randomness,
\begin{equation}
    \sum_{\lambda}q(\lambda)
    \operatorname{Tr}
    \left(\rho_x^\lambda\Pi_d^\lambda\right)
    \geq 1-\varepsilon
    \qquad \forall x,
    \label{eq:average_leakage}
\end{equation}
which allows realizations with larger leakage to be compensated by others with smaller leakage. As in the mean-observable-constrained framework of Ref.~\cite{van2017semi}, the average condition is closer to what can be directly characterized from the observed ensemble, without requiring access to each hidden variable $\lambda$, which motivates its use throughout this work. An important remark is that Eq.~\eqref{eq:average_leakage} allows for particular strategies to exceed the leakage bound for a specific $\lambda$, as long as the average remains bounded by what is observed in the experiment.

The classical counterpart, illustrated in Fig.~\ref{fig:PAMdiagram}(b), is
\begin{equation}
    \sum_{\lambda}q(\lambda)
    \sum_{m\in\mathcal{M}_d^\lambda}
    p(m|x,\lambda)
    \geq 1-\varepsilon
    \qquad \forall x,
    \label{eq:classical_leakage}
\end{equation}
where $\mathcal{M}_d^\lambda$ is the designated alphabet for a given $\lambda$, common to all preparation inputs $x$, with $|\mathcal{M}_d^\lambda|\leq d$ for every $\lambda$. As in the quantum case, particular realizations may have a larger leakage, as long as the average over $\lambda$ remains bounded.

Linear dimension witnesses provide experimentally accessible tests of communication constraints. For four preparations, $x\in\{0,1,2,3\}$, and two binary measurements, $y\in\{0,1\}$, we consider the witness~\cite{DavidePAM}
\begin{equation}
    \begin{aligned}
        S_4
        &=
        E_{00}+E_{01}+E_{10}-E_{11}\\
        &\quad
        -E_{20}+E_{21}-E_{30}-E_{31},
    \end{aligned}
    \label{eq:witness_S4}
\end{equation}
where $E_{xy}=p(0|x,y)-p(1|x,y)$. For a strict bit constraint, its classical bound is $S_4^{\mathrm C}=4$, while the maximal qubit value is $S_4^{\mathrm Q}=4\sqrt{2}$. Other witnesses and scenarios are considered in Appendix~\ref{app:main_scenarios}, together with their results.

Under the average leakage assumption of Eqs.~\eqref{eq:average_leakage} and \eqref{eq:classical_leakage}, these bounds change, since the communicated system is no longer fully restricted to dimension $d$, and become
\begin{equation}
    S_4^{\mathrm C}(\varepsilon)
    =\min\{4+8\varepsilon,8\},
    \label{eq:S4_C_ep}
\end{equation}
and
\begin{equation}
    S_4^{\mathrm Q}(\varepsilon)
    =
    \begin{cases}
        4\sqrt{2+8\varepsilon-8\varepsilon^2},
        & 0\leq\varepsilon\leq\frac12,\\
        8,
        & \varepsilon\geq\frac12,
    \end{cases}
    \label{eq:S4_Q_ep}
\end{equation}
where we derive the classical bound in Appendix~\ref{app:classical}, and the quantum bound follows from the optimal almost-qubit random access code in Ref.~\cite{Pauwels2022}, whose success probability relates to our witness by $p_{\mathrm{RAC}}=1/2+S_4/16$. Both bounds recover the usual bit and qubit values at $\varepsilon=0$ and reach the algebraic maximum $S_4=8$ at $\varepsilon=1/2$.

\section{Average randomness is more robust to dimension leakage}\label{sec:average_randomness}
In the original work introducing the almost-qudit framework~\cite{Pauwels2022}, as well as in several other semi-device-independent works tailored to randomness certification~\cite{Li2012,Mironowicz2014,van2017semi,Tavakoli2021,Tavakoli2022Info,Roch2025,Roch2025EAT}, it is commonly assumed that a fixed-input pair $(x_0,y_0)$ is used for randomness generation, while the remaining inputs are mainly used to estimate the value of the witness. In this setting, for a decomposition of the observed behavior into strategies $\lambda$ known to the adversary, we define the guessing probability for the fixed-input pair as
\begin{equation}
P_g(x_0,y_0)=\sum_{\lambda}q(\lambda)\max_b p(b|x_0,y_0,\lambda),
\label{eq:P_fixed}
\end{equation}
with the corresponding certified randomness quantified by the min-entropy $H_{\min}=-\log_2 P_g(x_0,y_0)$. One of the main reasons to consider such a type of estimator is that the witness and the choice of $(x_0,y_0)$ can be tailored such that the randomness is concentrated on this particular input pair, therefore allowing more randomness to be extracted from it than from other inputs, or from an average over all of them.

We consider, however, another estimator that remains less explored, although some instances of it have already appeared in the literature~\cite{Li2015,Mironowicz2016,alves2026semi}. We call it the average uniform guessing probability, defined as
\begin{equation}
\overline{P_g}=\frac{1}{|X||Y|}\sum_{x,y}\sum_{\lambda}q(\lambda)\max_b p(b|x,y,\lambda),
\label{eq:P_avg}
\end{equation}
which captures the guessing probability when all input pairs are used for randomness generation and are sampled uniformly, with the corresponding min-entropy given by $H_{\min}^{\mathrm{avg}}=-\log_2\overline{P_g}$. Notice that, for uniformly sampled inputs, this quantity is simply the average of the fixed-input guessing probabilities over all pairs $(x,y)$.

The maximizations over $b$ in Eqs.~\eqref{eq:P_fixed} and~\eqref{eq:P_avg} are not linear, since the most likely outcome may depend on the adversary's classical information $\lambda$ and, for the average estimator, also on the input pair $(x,y)$. To remove this maximization, we group together all realizations $\lambda$ that lead to the same deterministic guess. For the fixed-input estimator, a branch is labelled by a single guess $g\in\{0,1\}$ for $(x_0,y_0)$, while for the average estimator it is labelled by a complete table $\boldsymbol{g}=\{g_{xy}\}_{x,y}$. Denoting the total probability of each branch by $q_g$ and defining the subnormalized correlator $E_{xy}^g=q_g\sum_{b=0}^{1}(-1)^b p(b|x,y,g)$, the guessing probabilities can be written linearly as
\begin{align}
P_g(x_0,y_0)
&=
\sum_{g=0}^{1}
\frac{q_g+(-1)^gE_{x_0y_0}^g}{2},
\label{eq:P_fixed_branch}\\
\overline{P_g}
&=
\sum_{\boldsymbol{g}}
\left[
\frac{q_{\boldsymbol{g}}}{2}
+
\frac{1}{2|X||Y|}
\sum_{x,y}
(-1)^{g_{xy}}E_{xy}^{\boldsymbol{g}}
\right],
\label{eq:P_avg_branch}
\end{align}
which can now be used as proper linear objective functions in optimization problems.

The goal now is to seek for the largest guessing probability for a given witness value $s=\sum_{x,y}C_{xy}E_{xy}$ and leakage $\varepsilon$.  To upper bound these values, we use the SDP hierarchy introduced for almost qudits in Ref.~\cite{Pauwels2022}. In their construction, the preparations can be taken to be pure and are therefore represented by rank-one projectors $R_x$, while the measurement effects can be represented by projectors $M_{b|y}$ through a Naimark dilation. The $d$-dimensional subspace is represented by a rank-$d$ projector $V$. The almost-$d$ condition for a normalized preparation is then $\operatorname{Tr}(R_xV)\geq1-\varepsilon$. 

Notice now that, because Eqs.~\eqref{eq:P_fixed_branch} and \eqref{eq:P_avg_branch} optimize over guessing branches weighted by $q_g$, it is convenient to work directly with subnormalized quantities, absorbing this weight into a weighted trace $\tau_g(A)=q_g\operatorname{Tr}(A)$. This gives, in particular, $\tau_g(R_x)=q_g$ and $\tau_g(V)=dq_g$. For the leakage assumption, we can either assume Eq.~\eqref{eq:branchwise_leakage} or Eq.~\eqref{eq:average_leakage}. We use the latter, which in this notation reads
\begin{equation}
    \sum_g\tau_g(R_xV)\geq1-\varepsilon
\end{equation}
for all $x$, as it imposes a weaker assumption on the adversary.

What we do next is to replace the unknown operator products appearing inside $\tau_g$ by moment matrices. For each branch we construct a moment matrix $\Gamma^g$ whose entries are $\Gamma^g_{u,v}=\tau_g(uv^\dagger)$, with $u$ and $v$ products of the operators $\{\mathbbm{1},V,\{R_x\}_x,\{M_{b|y}\}_{b,y}\}$ up to the chosen relaxation level. In this way, every quantity involving $\tau_g$ that appears in the optimization is simply an entry, or a linear combination of entries, of $\Gamma^g$; for instance, $\tau_g(R_x)=\Gamma^g_{\mathbbm{1},R_x}$, $\tau_g(R_xM_{b|y})=\Gamma^g_{R_x,M_{b|y}}$, and $\tau_g(R_xV)=\Gamma^g_{R_x,V}$. By construction, $\Gamma^g\succeq0$. The algebraic relations satisfied by the operators, such as $R_x^2=R_x$, $M_{b|y}^2=M_{b|y}$, $V^2=V$, and $\sum_bM_{b|y}=\mathbbm{1}$, together with cyclicity of the trace, translate into linear equalities between different entries of $\Gamma^g$. For the binary-outcome witnesses considered below, we write $M_y\equiv M_{0|y}$. Denoting by $G$ the guessing-probability objective of either Eq.~\eqref{eq:P_fixed} or Eq.~\eqref{eq:P_avg}, the resulting SDP is
\begin{align}
\underset{\{q_g,\Gamma^g\}}{\operatorname{maximize}}\quad
& G \label{eq:main_sdp}\\
\text{subject to}\quad
&\Gamma^g\succeq0,\qquad q_g\geq0,\qquad \sum_gq_g=1, \nonumber\\
&\sum_g\sum_{x,y}C_{xy}E_{xy}^g=s, \nonumber\\
&E_{xy}^g=2\tau_g(R_xM_y)-q_g, \nonumber\\
&\tau_g(R_x)=q_g,\qquad \tau_g(V)=dq_g, \nonumber\\
&\sum_g\tau_g(R_xV)\geq1-\varepsilon, \nonumber\\
&0\leq\tau_g(R_xM_y)\leq q_g,
\nonumber\\
&0\le \tau_g(R_xV)\le q_g, \qquad \forall x,y,g, \nonumber
\end{align}
where all terms $\tau_g(\cdot)$ are understood as linear entries of $\Gamma^g$, while the weights $q_g$ are fixed by $\tau_g(R_x)=q_g$. For Eq.~\eqref{eq:P_fixed}, $g$ denotes a single binary guess, while for Eq.~\eqref{eq:P_avg} it denotes a complete table of guesses $\boldsymbol{g}$. Since the moment construction relaxes the original quantum problem, its optimum $G^\star$ upper bounds the guessing probability and therefore gives the certified lower bound $H_{\min}^{\star}=-\log_2G^\star$.

\begin{figure}[t]
    \centering
    \includegraphics{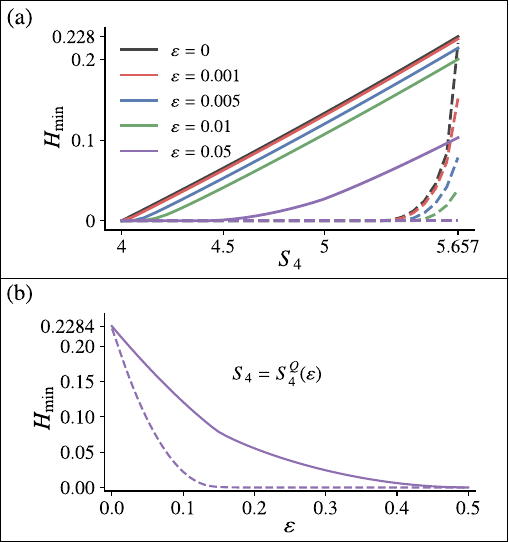}
    \caption{\textbf{Average versus fixed-input randomness under dimension leakage.}
    (a) Certified min-entropy for the $S_4$ witness as a function of its observed value and for different leakage parameters $\varepsilon$. Solid curves correspond to the average uniform guessing probability and dashed curves to the fixed-input guessing probability with $(x_0=0,y_0=0)$.
    (b) Certified min-entropy evaluated at the maximal almost-qubit value $S_4^{Q}(\varepsilon)$ as a function of $\varepsilon$. The solid purple curve shows the average case, while the dashed purple curve shows the fixed-input one.}
    \label{fig:S4-avg-fixed}
\end{figure}

We next compare the two guessing probabilities for the witnesses considered here. Since the same qualitative behavior is observed in all cases, we focus in the main text on $S_4$ and leave the results for other scenarios witnesses to the Appendix~\ref{app:main_scenarios}. All results are obtained at the second level of the moment hierarchy. For a strict dimension bound, $\varepsilon=0$, the two estimators behave similarly close to the maximal quantum violation, as shown in Fig.~\ref{fig:S4-avg-fixed}(a). As soon as dimension leakage is introduced, while the fixed-input estimator only becomes positive for sufficiently large values of $S_4$, the average estimator certifies randomness already for much smaller violations and decreases much more smoothly as $\varepsilon$ is increased. For $\varepsilon=0.05$, the fixed-input estimator does not certify randomness over the range shown, while the average estimator still gives a nonzero min-entropy for several values of $S_4$.

We must emphasize, however, that there is a trade-off between the two estimators. A fixed-input pair can be chosen such that the witness certifies particularly large randomness for that pair, and can therefore outperform the average estimator under a strict dimension assumption (see the Appendix~\ref{app:main_scenarios} result for the $S_3$ and $S_{331}$ witnesses). On the other hand, the average estimator is less affected by leakage because increasing the predictability of only the selected pair is no longer enough. Since all input pairs contribute to the guessing probability, the adversary needs to improve the predictability over a larger part of the input set, which gives an intuition for the slower degradation seen in Fig.~\ref{fig:S4-avg-fixed}(b), where we evaluate the min-entropy of both cases for the maximal almost-qubit value $S_4^{Q}(\varepsilon)$ for each leakage.

Another important point to notice in Fig.~\ref{fig:S4-avg-fixed}(b) is that, for the average case, randomness can in principle still be certified throughout the region where the quantum and classical bounds in Eqs.\eqref{eq:S4_C_ep} and \eqref{eq:S4_Q_ep} remain separated, up to the point where both reach the algebraic maximum $S_4=8$ at $\varepsilon = 1/2$. For the fixed-input case, the certification disappears much earlier.

\begin{figure*}[t!]
\includegraphics{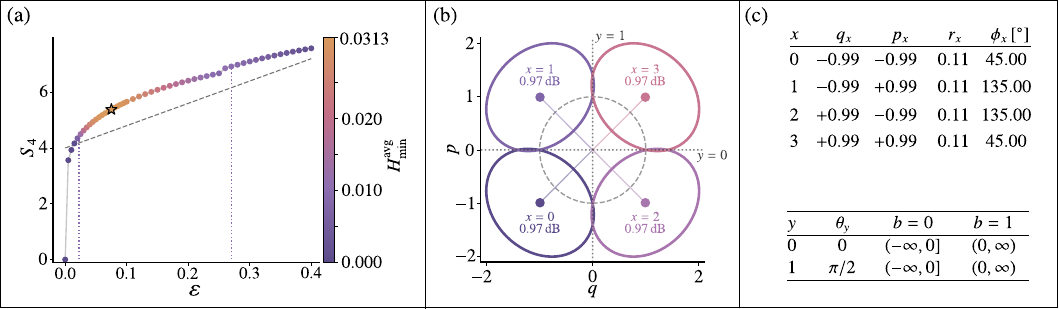}
\caption{
\textbf{Gaussian preparations with homodyne detection.} (a) Optimized $S_4$ as a function of the common leakage $\varepsilon$, with color indicating the lower bound on $H_{\min}^{\mathrm{avg}}$ and the dashed line the classical almost-qubit bound. The dotted vertical lines delimit the region where  $H_{\min}^{\mathrm{avg}}\geq 0.01$, and the star marks the optimal value, at $\varepsilon=0.075$, $S_4\simeq5.379$, and $H_{\min}^{\mathrm{avg}}\geq0.0313$ bits. (b) Phase-space representation of the preparations: dots show the displacements, solid curves the $1\sigma$ contours, the dashed circle the vacuum $1\sigma$ contour, and the dotted axes the two homodyne quadratures. (c) Preparation and measurement parameters for the same realization, with $\bar n_x=0$ for all preparations.}
\label{fig:gauss-homodyne}
\end{figure*}

\section{Dimension leakage as a resource for continuous-variable implementations}\label{sec:cv_implementations}
To our knowledge, a single-mode fully Gaussian setup (Gaussian preparations and measurements) has not been shown to detect nonclassicality---and therefore certify randomness---in the hard-truncated $d$-dimensional PM scenario. Allowing leakage makes violations of the classical bound in Eq.~\eqref{eq:S4_C_ep} possible, but at the same time weakens the randomness bounds. Here we show that the increased robustness of the average estimator allows this tradeoff to be balanced, leading to positive randomness from a fully Gaussian setup.

We also investigate other CV implementations, including Gaussian preparations with displaced photodetection and non-Gaussian preparations with homodyne detection. The former allows us to assess whether non-Gaussian measurements can enhance randomness certification, while the latter explores the role of non-Gaussian preparations, considering both unrestricted states and experimentally implementable superpositions of squeezed cat states.

\subsection{Gaussian preparations with homodyne detection}
\label{subsec:fullgaussian}

We first consider a fully Gaussian implementation, with single-mode Gaussian preparations followed by homodyne detection. Following the phase-space description of Ref.~\cite{Weedbrook2012}, we use quadratures satisfying $[\hat q,\hat p]=2i$, such that the vacuum covariance matrix is the identity. A single-mode Gaussian preparation $\rho_x$ is completely characterized by its displacement vector $\boldsymbol d_x=(q_x, p_x)^T$ and its covariance matrix $V_x$. We parametrize it as
\begin{equation}
    V_x=
    (2\bar n_x+1)
    R(\phi_x)
    \begin{pmatrix}
        e^{-2r_x} & 0\\
        0 & e^{2r_x}
    \end{pmatrix}
    R^{T}(\phi_x),
    \label{eq:gaussian_covariance}
\end{equation}
where $\bar n_x\geq0$ is the thermal occupation, $r_x\geq0$ is the squeezing parameter, $\phi_x$ specifies the squeezing direction, and $R(\phi)$ denotes the usual rotation matrix by an angle $\phi$. Pure Gaussian states are recovered for $\bar n_x=0$.

For measurement input $y$, homodyne detection measures the quadrature $\hat q_{\theta_y}=\hat q\cos\theta_y+\hat p\sin\theta_y$, where $\theta_y$ is the local-oscillator phase. For a Gaussian preparation $\rho_x$, the continuous homodyne outcome $z\in\mathbb{R}$ follows a normal distribution~\cite{Brask2022} with probability density
\begin{equation}
    f_{xy}(z)
    =
    \frac{1}{\sqrt{2\pi\sigma_{xy}^2}}
    \exp\left[
        -\frac{(z-\mu_{xy})^2}{2\sigma_{xy}^2}
    \right],
    \label{eq:gaussian_homodyne_distribution}
\end{equation}
with mean $\mu_{xy}=\boldsymbol u_y^T\boldsymbol d_x$ and variance $\sigma_{xy}^2=\boldsymbol u_y^T V_x\boldsymbol u_y$, where $\boldsymbol u_y=(\cos\theta_y,\sin\theta_y)^T$ specifies the measured quadrature. The probability of obtaining an outcome in a measurable interval $\mathcal R\subseteq\mathbb{R}$ is therefore $\Pr(z\in \mathcal R|x,y)=\int_{\mathcal R} f_{xy}(z)\,dz$, with $\int_{-\infty}^{\infty} f_{xy}(z)\,dz=1$. To obtain a binary behavior, we assign the interval $\mathcal R_y=[a_y,b_y]$ to outcome $b=0$ and its complement $\mathbb{R}\setminus\mathcal R_y$ to $b=1$. The resulting probabilities can then be evaluated analytically as
\begin{align}
    p(0|x,y)
    &=
    \Phi\left(\frac{b_y-\mu_{xy}}{\sigma_{xy}}\right)
    -
    \Phi\left(\frac{a_y-\mu_{xy}}{\sigma_{xy}}\right),
    \label{eq:gaussian_homodyne_behavior}\\
    p(1|x,y)
    &=1-p(0|x,y),
\end{align}
where $\Phi(t)=1/{\sqrt{2\pi}}\int_{-\infty}^{t}e^{-u^2/2}\,du$ denotes the cumulative distribution function of a standard normal distribution. The correlators entering $S_4$ (and also the other inequalities considered in Appendix~\ref{app:main_scenarios}) are therefore $E_{xy}=2p(0|x,y)-1$.

For the almost-qubit condition, the leakage of preparation $x$ is the population outside the first two Fock levels, $\varepsilon_x=1-\langle0|\rho_x|0\rangle-\langle1|\rho_x|1\rangle$. For a single-mode Gaussian state, the Fock-state populations can be obtained analytically~\cite{Dodonov1994}. Since Ref.~\cite{Dodonov1994} uses a different quadrature normalization and covariance-matrix convention from the one adopted here, we provide in Appendix~\ref{app:gaussian_fock} a short derivation directly in our notation. Defining $A_x=V_x+\mathbbm{1}$ and $J_x=A_x^{-1}$, the vacuum and single-photon populations are
\begin{align}
    p_{0,x}
    &=
    \langle0|\rho_x|0\rangle
    =
    \frac{2\exp[-\boldsymbol d_x^T J_x\boldsymbol d_x/2]}
    {\sqrt{\det A_x}},
    \\
    p_{1,x}
    &=
    \langle1|\rho_x|1\rangle
    =
    p_{0,x}
    \left(
        1-\operatorname{Tr}J_x
        +\boldsymbol d_x^T J_x^2\boldsymbol d_x
    \right),
\end{align}
such that $\varepsilon_x=1-p_{0,x}-p_{1,x}$.

\begin{figure*}[t!]
\includegraphics{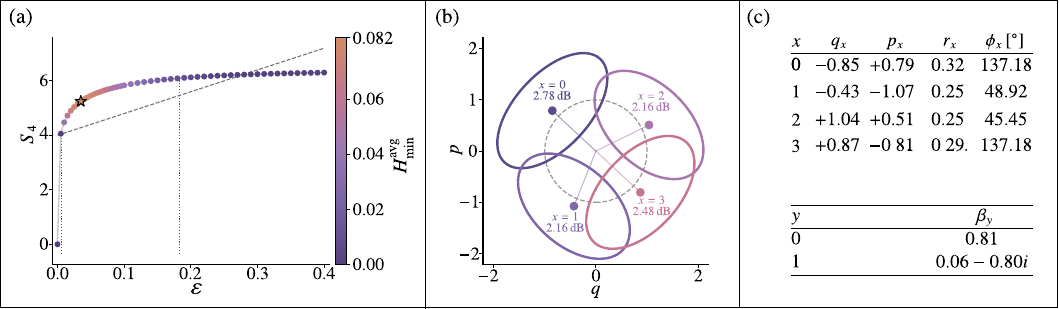}
\caption{\textbf{Gaussian preparations with displaced photodetection.}
(a) Certified randomness under dimension leakage. The dashed curve gives the classical bound, the dotted vertical lines delimit the region with $H_{\min}\geq0.01$, and the star marks the optimal randomness point, at $\varepsilon=0.035$, $S_4\simeq5.245$, and $H_{\min}^{\mathrm{avg}}\geq0.082$ bits.
(b) Phase-space representation of the optimized Gaussian preparations.
(c) Preparation and measurement parameters at $\varepsilon=0.035$, with $\bar n_x=0$ for all preparations.}
\label{fig:gauss-displacement}
\end{figure*}

We use these expressions to numerically search for large values of $S_4$ at a fixed leakage. Let $\boldsymbol\xi=\{q_x,p_x,r_x,\phi_x,\bar n_x,\theta_y,a_y,b_y\}_{x,y}$. For each chosen value of $\varepsilon$, we consider the nonlinear optimization problem
\begin{align}
    \underset{\boldsymbol\xi}{\operatorname{maximize}}
    \quad & S_4(\boldsymbol\xi)
     \label{eq:gaussian_homodyne_optimization}\\
    \text{subject to}\quad
    &\varepsilon_x(\boldsymbol\xi)=\varepsilon,
    \qquad x=0,1,2,3,
    \nonumber
\end{align}
which we solve numerically using global differential-evolution searches. The resulting points are not guaranteed to give the global Gaussian maximum and we therefore interpret them only as optimized feasible points. This is sufficient for our purpose, since any such point already shows that the corresponding pair $(s,\varepsilon)$, with $s=S_4(\boldsymbol\xi)$, can be physically achieved using Gaussian preparations and homodyne measurements.

For each optimized pair $(s,\varepsilon)$, we then solve the SDP in Eq.~\eqref{eq:main_sdp} and obtain a lower bound on $H_{\min}^{\mathrm{avg}}$. The result of this procedure is shown in Fig.~\ref{fig:gauss-homodyne}(a). As the allowed leakage increases, the optimized Gaussian value of $S_4$ also increases, eventually exceeding the corresponding classical almost-qubit bound. However, the certified randomness does not follow the same monotonic behavior. Instead, it is maximized at an intermediate leakage, showing the tradeoff between achieving a larger violation and simultaneously weakening the dimension assumption available to the adversary. The best point found along this curve, marked by the star, occurs at $\varepsilon=0.075$, for which $S_4=5.38$ and $H_{\min}^{\mathrm{avg}}\geq0.0313$. Notice that, although larger values of $S_4$ can be reached by increasing $\varepsilon$, they lead to less certified randomness.

The Gaussian implementation is shown in Fig.~\ref{fig:gauss-homodyne}(b), with the parameters listed in Fig.~\ref{fig:gauss-homodyne}(c). The four preparations are arranged in the four quadrants of phase space, reproducing the sign pattern that underlies the $2\!\to\!1$ QRAC square~\cite{Ambainis2002, Ambainis2009QRAC}. The two homodyne settings measure the \(q\) and \(p\) quadratures, with the sign of each outcome determining the binary outcome $b$. It is therefore not exactly the usual QRAC square, but rather its continuous-variable analogue, now implemented with displaced weakly squeezed Gaussian states. The squeezing is also small, \(r\simeq0.11\) (\(0.97\,\mathrm{dB}\)), making the strategy experimentally feasible. We therefore close this section by showing that a fully Gaussian implementation can certify randomness when a controlled amount of dimension leakage is allowed.

\subsection{Gaussian preparations with displaced photodetection}

We next keep the same class of single-mode Gaussian preparations, but replace homodyne detection by a displaced on--off photodetection measurement. Such schemes consist of applying a phase-space displacement followed by photon detection~\cite{banaszek1996,BraskChaves2012,chaves2018causal}. Although the displacement is a Gaussian operation, photon counting is a non-Gaussian measurement.

For an ideal on--off detector, the no-click and click events are described by $\Pi_{\mathrm{nc}}=|0\rangle\langle0|$ and $\Pi_{\mathrm{c}}=\mathbbm{1}-|0\rangle\langle0|$, respectively. If the measurement input $y$ specifies a displacement $D(\beta_y)$ before detection, the effective no-click operator acting on the incoming state is
\begin{equation}
    M_{1|y}
    =
    D^\dagger(\beta_y) \, \Pi_{\mathrm{nc}} \, D(\beta_y)
    =
    |-\beta_y\rangle\langle-\beta_y|,
\end{equation}
where the sign of the displacement can be absorbed into its definition, such that we write simply $M_{1|y}=|\beta_y\rangle\langle\beta_y|$ and $M_{0|y}=\mathbbm{1}-|\beta_y\rangle\langle\beta_y|$. Notice that we use the convention $b=1$ for no-click and $b=0$ for click.

The corresponding no-click probability is therefore the overlap of the preparation with a coherent state (see Eq.~\eqref{eq:gaussian_coherent_overlap} in Appendix~\ref{app:gaussian_fock}),
\begin{equation}
\begin{aligned}
    p(1|x,y)
    &=
    \langle\beta_y|\rho_x|\beta_y\rangle
    \\
    &=
    \frac{2}{\sqrt{\det A_x}}
    \exp\left[
        -\frac12
        (\boldsymbol d_{\beta_y}-\boldsymbol d_x)^T
        J_x
        (\boldsymbol d_{\beta_y}-\boldsymbol d_x)
    \right],
\end{aligned}
\label{eq:displaced_photodetection}
\end{equation}
where $A_x=V_x+\mathbbm{1}$, $J_x=A_x^{-1}$, $\boldsymbol d_{\beta_y}=(2\operatorname{Re}\beta_y,2\operatorname{Im}\beta_y)^T$. The correlators can then be obtained as $E_{xy} = 1 - 2p(1|x,y)$.

\begin{figure*}[t]
    \centering
    \includegraphics[width=\textwidth]{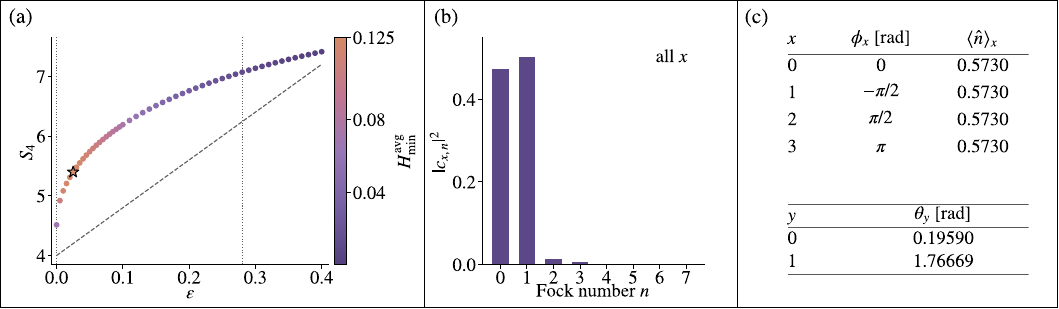}
    \caption{\textbf{Unrestricted non-Gaussian states with homodyne detection.}
    (a) Certified randomness under dimension leakage. The dashed curve gives the classical bound, the dotted vertical lines delimit the region with $H^{\mathrm{avg}}_{\min}\geq0.01$, and the star marks the largest-randomness point, at $\varepsilon=0.025$, $S_4\simeq5.398$ and $H^{\mathrm{avg}}_{\min} \geq 0.125$ bits.
    (b) Fock-state populations $|c_{x,n}|^2$ at the starred point,
    common to all four preparations.
    (c) Preparation phases $\phi_x$, mean photon numbers
    $\langle\hat n\rangle_x$, and homodyne phases $\theta_y$ for
    the same realization. Up to global phases, the four preparations
    satisfy $|\psi_x\rangle=e^{i\phi_x\hat n}|\psi_0\rangle$ to
    numerical precision, with
    $|\psi_0\rangle \simeq
      -0.6877|0\rangle
      +(0.3939+0.5889i)|1\rangle
      +(0.0452-0.1094i)|2\rangle
      -(0.0821-0.0164i)|3\rangle
      +(0.0252+0.0252i)|4\rangle
      -(0.0055-0.0280i)|5\rangle
      +(0.0215-0.0090i)|6\rangle
      -(0.0308+0.0204i)|7\rangle$.}
    \label{fig:nongaus-homodyne}
\end{figure*}

We again solve Eq.~\eqref{eq:gaussian_homodyne_optimization}, now with $\boldsymbol\xi=\{q_x,p_x,r_x,\phi_x,\bar n_x,\beta_y\}_{x,y}$, such that the optimization runs over both the Gaussian preparation parameters and the complex displacement amplitudes $\beta_y$. For each optimized pair $(s,\varepsilon)$, we then solve the SDP in Eq.~\eqref{eq:main_sdp}. The results are shown in Fig.~\ref{fig:gauss-displacement}, following the same structure as in the previous subsection. Notice that the certified randomness in Fig.~\ref{fig:gauss-displacement}(a) is improved when displaced photodetection is used instead of homodyne detection for the same class of Gaussian preparations. The phase-space representation of the preparations is shown in Fig.~\ref{fig:gauss-displacement}(b), while the corresponding preparation and measurement parameters are listed in Fig.~\ref{fig:gauss-displacement}(c).

\subsection{Non-gaussian preparations with homodyne detection}
\subsubsection{Unrestricted states with a Fock cutoff}\label{unrestricted-states}
To study the role of the preparations, we retain the binary homodyne measurements of Sec.~IV\,A and consider arbitrary pure states. For numerical tractability, we introduce a cutoff at the first eight Fock levels, writing $|\psi_x\rangle=\sum_{n=0}^{7}c_{x,n}|n\rangle$, with $\sum_{n=0}^{7}|c_{x,n}|^2=1$. For a target qubit encoded in $|0\rangle$ and $|1\rangle$, we impose a common leakage for all preparations, $\varepsilon_x=1-|c_{x,0}|^2-|c_{x,1}|^2=\varepsilon$.

Since we introduced a Fock cutoff, we can follow the homodyne construction of Ref.~\cite{alves2026semi}, now extended to the first eight Fock levels. For each measurement input $y$, the measured quadrature $\hat q_{\theta_y}$ is binned into an interval $[a_y,b_y]$ and its complement, with the corresponding measurement operators represented as finite-dimensional matrices in the truncated Fock basis. For fixed measurement settings and outcome assignments, the witness can be written as
\begin{equation}
    S_4=\sum_x \boldsymbol{c}_x^\dagger W_x\boldsymbol{c}_x,
    \qquad
    W_x=\sum_y C_{xy}(2M_y-\mathbbm{1}),
\end{equation}
where $\boldsymbol{c}_x=(c_{x,0},\ldots,c_{x,7})^T$, $C_{xy}$ are the coefficients of Eq.~\eqref{eq:witness_S4}, and $M_y$ is the measurement operator associated with outcome $b=0$. Since each term depends only on one preparation, the Fock amplitudes can be optimized independently for each $x$, subject to the normalization and leakage constraints. We then optimize the homodyne phases and interval boundaries using differential evolution followed by local refinement, considering both outcome assignments. For each $\varepsilon$, the resulting value of $S_4$ is used in the SDP of Eq.~\eqref{eq:main_sdp} to certify randomness under the average leakage assumption of Eq.~\eqref{eq:average_leakage}.

Fig.~\ref{fig:nongaus-homodyne}(a) shows that the largest lower bound on the average min-entropy found in the scan occurs at $\varepsilon=0.025$, where $S_4\simeq5.398$ and the bound is approximately $0.125$ bits per round. At the same point, the fixed-input bound for $(x_0,y_0)=(0,0)$ is numerically zero. For comparison, the best Gaussian realization found with homodyne detection gives $0.0313$ bits at $\varepsilon=0.075$ (Fig.~\ref{fig:gauss-homodyne}(a)). The non-Gaussian realization thus gives about four times the certified average min-entropy at one third of the leakage.

Another noteworthy point in Fig.~\ref{fig:nongaus-homodyne}(a) is that, for $\varepsilon=0$, we recover the homodyne result of Ref.~\cite{alves2026semi}, where arbitrary preparations within the first two Fock levels were considered. Interestingly---and as suggested by Ref.~\cite{alves2026semi}---even without dimensional leakage, we already certify a positive average min-entropy of approximately $0.067$ bits (which can be enhanced to $0.125$ bits when a small leakage is allowed).

At the starred point, the four independently optimized states are, up to global phases and numerical precision, related by $|\psi_x\rangle=e^{i\phi_x\hat n}|\psi_0\rangle$, with $\phi_x=0,-\pi/2,\pi/2,\pi$ for $x=0,1,2,3$, respectively (where $|\psi_0\rangle$ is given in the caption of Fig.~\ref{fig:nongaus-homodyne}). These rotations preserve the Fock populations, so the distribution in Fig.~\ref{fig:nongaus-homodyne}(b) applies to all four states. The vacuum and single-photon populations are approximately $0.473$ and $0.502$, leaving $0.025$ in the higher Fock levels. The corresponding homodyne phases in Fig.~\ref{fig:nongaus-homodyne}(c) differ by $\pi/2$ to numerical precision, and the optimized intervals effectively bin each quadrature by its sign. As in the Gaussian case, the preparations exhibit a QRAC-like square symmetry, now generated by phase rotations of a single non-Gaussian state.

\subsubsection{Superpositions of squeezed cat states}

Although the unrestricted eight-level optimization above establishes the performance achievable within a truncated Fock space, preparing arbitrary superpositions of Fock states remains experimentally challenging. Motivated by the rotational symmetry e parity structure observed in the previous section, we now restrict the preparations to a physically structured family based on superpositions of even and odd squeezed optical cat states~\cite{ourjoumtsev2007generation,sychev2017enlargement,schlegel2022quantum}.

Similar to the case of unrestricted states with a Fock cutoff, we generate the four preparations from a single seed state $|\psi_0\rangle$ through the phase-space rotations $|\psi_x\rangle = e^{i\phi_x\hat{n}}|\psi_0\rangle$, with $\phi_x = 0, -\pi/2, \pi/2, \pi$ for $x = 0, 1, 2, 3$, respectively. Because the sign-binned homodyne operators only couple Fock states of opposite parity, a state with definite photon-number parity yields vanishing correlators. But this time we construct $|\psi_0\rangle$ as a coherent superposition of orthogonal even ($+$) and odd ($-$) squeezed cat states,
\begin{equation}
    |\psi_0\rangle = \sqrt{w_+}\,|\psi_+\rangle + e^{i\chi}\sqrt{1 - w_+}\,|\psi_-\rangle,
    \label{eq:seed_squeezed_cat}
\end{equation}
with $0 \le w_+ \le 1$, relative phase $\chi \in [-\pi, \pi]$, and
\begin{equation}
    |\psi_\pm\rangle = \mathcal{N}_\pm \hat{S}(r_\pm)\left(|\alpha_\pm\rangle \pm |-\alpha_\pm\rangle\right),
    \label{eq:squeezed_cat_components}
\end{equation}
where $\hat{S}(r_\pm) = \exp\big[\frac{r_\pm}{2}(\hat{a}^2 - \hat{a}^{\dagger 2})\big]$ is the single-mode squeezing operator with real squeezing parameter $r_\pm$, $|\pm\alpha_\pm\rangle$ are coherent states with amplitude $\alpha_\pm \ge 0$, and $\mathcal{N}_\pm$ are normalization constants. This formulation is mainly inspired in Ref. \cite{schlegel2022quantum}.

Since $|\psi_+\rangle$ and $|\psi_-\rangle$ have support exclusively on even and odd Fock numbers, respectively, the vacuum and single-photon populations of each preparation are given by $\langle 0|\rho_x|0\rangle = w_+ |\langle 0|\psi_+\rangle|^2$ and $\langle 1|\rho_x|1\rangle = (1 - w_+) |\langle 1|\psi_-\rangle|^2$. Consequently, denoting the component leakages by $\epsilon_+ = 1 - |\langle 0|\psi_+\rangle|^2$ and $\epsilon_- = 1 - |\langle 1|\psi_-\rangle|^2$, the total leakage of $|\psi_0\rangle$ is the convex combination
\begin{equation}
    \epsilon_x = w_+ \epsilon_+ + (1 - w_+) \epsilon_- = \epsilon.
    \label{eq:cat_leakage}
\end{equation}

For any choice of parameters $(\alpha_\pm, r_\pm)$ such that $\epsilon$ lies between $\epsilon_+$ and $\epsilon_-$, the weight $w_+ = (\epsilon - \epsilon_-)/(\epsilon_+ - \epsilon_-)$ is uniquely fixed by Eq.~\eqref{eq:cat_leakage}, ensuring that the exact leakage constraint is satisfied analytically. At $\epsilon = 0$, taking $\alpha_\pm, r_\pm \to 0$ reduces $|\psi_+\rangle$ and $|\psi_-\rangle$ to $|0\rangle$ and $|1\rangle$, naturally recovering the strict-qubit preparations. For $\epsilon > 0$, we numerically optimize $S_4$ over the state parameters $(\alpha_\pm, r_\pm, \chi)$ and the homodyne angles $\theta_y$ (with sign-binned outcomes) in a truncated Fock space, and insert each optimized pair $(S_4, \epsilon)$ into the SDP of Eq.~\eqref{eq:main_sdp} to lower bound $H_{\min}^{\mathrm{avg}}$.

\begin{figure}[t]
    \centering
    \includegraphics[width=\columnwidth]{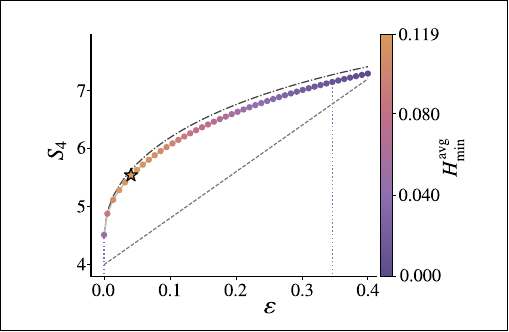}
    \caption{\textbf{Superpositions of squeezed cat states with homodyne detection.} Optimized $S_4$ as a function of the common leakage $\varepsilon$, with color indicating the lower bound on $H_{\min}^{\mathrm{avg}}$ and the gray dashed line the classical almost-qubit bound $S_4^C(\epsilon)$. The black dash-dotted curve shows the witness values obtained for unrestricted states with a cutoff at the first eight Fock levels, and the purple dotted vertical lines delimit the region with $H_{\min}^{\mathrm{avg}} \ge 0.01$. The star marks the optimal randomness point, at $\epsilon = 0.041$, $S_4 \simeq 5.59$, and $H_{\min}^{\mathrm{avg}} \ge 0.119$ bits.}
    \label{F-nonGaussHomodyne_squeezCatStates}
\end{figure}

The results are presented in Fig.~\ref{F-nonGaussHomodyne_squeezCatStates}. As shown in the figure, the optimized witness values for the squeezed cat states closely track those of the unrestricted Fock-state search (dash-dotted curve) across the entire leakage range. The largest certified randomness occurs at $\epsilon = 0.041$, where $S_4 \simeq 5.59$ and $H_{\min}^{\mathrm{avg}} \ge 0.119$ bits per round. Remarkably, this structured family recovers about $95\%$ of the maximum min-entropy certified by unrestricted eight-level Fock states ($0.125$ bits at $\epsilon = 0.025$ in Fig.~\ref{fig:nongaus-homodyne}(a)), while outperforming the Gaussian homodyne implementation ($0.0313$ bits at $\epsilon = 0.075$ in Fig.~\ref{fig:gauss-homodyne}(a)) by nearly a factor of four.

At the starred point, all four preparations share the same photon-number distribution and a mean photon number of $\langle\hat{n}\rangle_x \approx 0.604$. The vacuum and single-photon populations are approximately $0.453$ and $0.506$, in that order, while the leaked population ($\epsilon = 0.041$) remains almost entirely confined to the two- and three-photon levels ($n = 2, 3$), with negligible contributions for $n \ge 4$, a behavior similar to Fig.~\ref{fig:nongaus-homodyne}(b). On the measurement side, the optimal homodyne phases are given by the orthogonal settings $\theta_0 = \pi/4$ and $\theta_1 = 3\pi/4$, with outcome $b = 0$ assigned to the intervals $[0, \infty)$ and $(-\infty, 0]$, respectively, and $b = 1$ to their complements.

\section{Discussion}
\label{sec:conclusao}

We have investigated non-classicality and randomness certification against dimensional leakage in the almost-qubit prepare-and-measure scenario. Our results show that the choice of randomness estimator plays a crucial role. While leakage can severely compromise certification based on a fixed preparation-measurement pair, averaging the guessing probability over the full input ensemble can mitigate that. The intuition is that an adversary can exploit additional dimensions to increase the predictability of a selected input pair, whereas the average estimator accounts for predictability across all preparation and measurement inputs. 

We have further shown that this robustness can enable continuous-variable implementations using experimentally accessible resources. In particular, single-mode Gaussian states measured by homodyne detection can certify nonzero average randomness even with relatively low squeezing. Replacing homodyne detection with displaced on-off photodetection increases the certified randomness in the Gaussian setting considered, but introduces a non-Gaussian measurement.

Non-Gaussian preparations with homodyne detection yield even larger certified randomness. While unrestricted states within the first eight Fock levels provide the best results, experimentally motivated squeezed cat states achieve nearly the same certification, retaining about 95\% of the optimal min-entropy found in the eight-level search.

More broadly, our results suggest that dimensional leakage need not be viewed solely as an imperfection to be minimized. Rather, within a suitable certification framework, the higher-dimensional structure of physical systems can be exploited to enable randomness certification in continuous-variable implementations.

Several directions for future work remain open. An important next step is to assess the approach under finite statistics and realistic experimental imperfections, including optical losses, detector inefficiencies, and finite measurement resolution, while developing methods to bound leakage from observed data. Generalizing the average guessing probability to nonuniform input distributions could help optimize certified randomness and clarify its trade-off with fixed-input certification. Further extensions to alternative dimension witnesses, larger input and output alphabets, and multimode optical fields may broaden the range of practical implementations.

\begin{acknowledgments}
We acknowledge funding the Brazilian National Council for Scientific and Technological Development (CNPq; Grants No. 403181/2024-0 and 301687/2025-0), the National Institute of Science and Technology for Applied Quantum Computing through CNPq Process No. 408884/2024-0, and the Financiadora de Estudos e Projetos (Grant No. 1699/24 IIF-FINEP). This work was also partially funded through the project “Design e arquiteturas de protocolos CV-QKD” supported by the EMBRAPII CIMATEC Competence Center in Quantum Technologies-Quantum Industrial Innovation (QuIIN), with financial resources from the PPI IoT/Manufatura 4.0 of the MCTI Grant No. 053/2023, signed with EMBRAPII. M.A. acknowledges financial support from the Center for Quantum-Enabled Computing (C4QEC) project, carried out within the International Research Agendas programme of the Foundation for Polish Science, co-financed by the European Union under the European Funds for Smart Economy 2021–2027 (FENG).

\end{acknowledgments}

\bibliographystyle{apsrev4-2}
\bibliography{ref}

\appendix
\section{Average versus fixed-input randomness for the $S_3$ and $S_{331}$ witnesses}
\label{app:main_scenarios}

The comparison between average and fixed-input randomness is extended here to two additional facet inequalities of the classical bit prepare-and-measure polytope. Both scenarios have three preparations, $x\in\{0,1,2\}$, binary outcomes $b\in\{0,1\}$, and target dimension $d=2$, with either two or three measurement inputs. As in the main text, the correlators are defined as $E_{xy}=p(0|x,y)-p(1|x,y)$.

For two measurement inputs, $y\in\{0,1\}$, the minimal dimension witness is
\begin{equation}
    S_3=E_{00}+E_{01}+E_{10}-E_{11}-E_{20}.
    \label{eq:app_S3}
\end{equation}
Its classical bit bound and maximal qubit value are $S_3^{C}(0)=3$ and $S_3^{Q}(0)=1+2\sqrt{2}$, respectively. For three measurement inputs, $y\in\{0,1,2\}$, the witness considered is
\begin{equation}
    S_{331}
    =E_{00}+E_{01}-E_{11}+E_{12}-E_{20}-E_{22},
    \label{eq:app_S331}
\end{equation}
with $S_{331}^{C}(0)=4$ and $S_{331}^{Q}(0)=3\sqrt{3}$. These bounds refer to the strict dimension constraint, $\varepsilon=0$.

For both witnesses, the average leakage condition of Eq.~(5) is imposed with $d=2$, separately for each preparation $x\in\{0,1,2\}$. The guessing probabilities are bounded using the SDP construction in Eq.~\eqref{eq:main_sdp}, with the witness coefficients replaced by those of Eqs.~\eqref{eq:app_S3} and~\eqref{eq:app_S331}.

For fixed-input generation, the selected pairs are $(x_0,y_0)=(2,1)$ for $S_3$ and $(x_0,y_0)=(0,2)$ for $S_{331}$. For average generation, all six or nine input pairs, respectively, are sampled uniformly, including those whose correlators do not appear in the witness. The quantities compared are $H_{\min}=-\log_2 P_g(x_0,y_0)$ and $H_{\min}^{\mathrm{avg}}=-\log_2\overline{P}_g$, with both guessing probabilities optimized under the same witness and leakage constraints.

\begin{figure}[t]
    \centering
    \includegraphics[width=\linewidth]
    {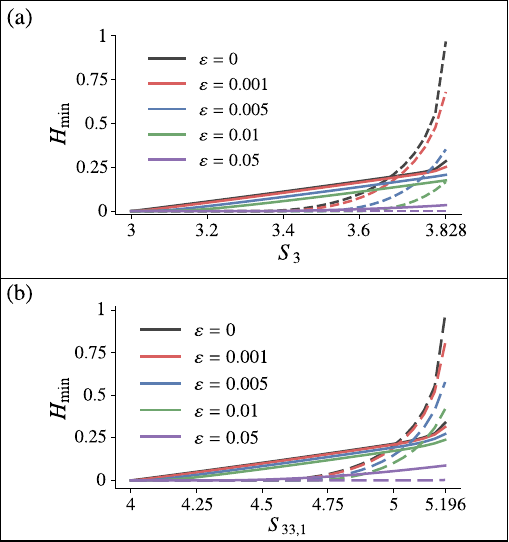}
    \caption{\textbf{Average versus fixed-input randomness
    under dimension leakage.}
    Certified min-entropy for the (a) $S_3$ and
    (b) $S_{331}$ witnesses as a function of their
    observed values and for different leakage
    parameters $\varepsilon$. Solid curves correspond
    to the average uniform guessing probability and
    dashed curves to the fixed-input guessing
    probability, with $(x_0,y_0)=(2,1)$ for $S_3$
    and $(x_0,y_0)=(0,2)$ for $S_{331}$.}
    \label{fig:S3-S331-avg-fixed}
\end{figure}

Fig.~\ref{fig:S3-S331-avg-fixed} shows the resulting certified lower bounds as functions of the observed witness value for $S_3$ in panel~(a) and $S_{331}$ in panel~(b). Under a strict dimension constraint, the selected fixed inputs yield more certified randomness close to the maximal qubit violation, whereas the average estimator certifies randomness over a broader range of witness values. Even at $\varepsilon=0$, there is therefore a trade-off between the amount of randomness certified near the quantum maximum and the range of violations for which certification is possible.

As the allowed dimensional leakage increases, the fixed-input bounds deteriorate rapidly, and larger witness values are required to obtain a positive certified min-entropy. The average bounds remain positive over a wider range and can exceed the fixed-input bounds even close to the ideal qubit maximum. For example, at $\varepsilon=0.05$ and $S_3\simeq3.8283$, the average estimator certifies approximately $0.034$ bits, whereas the fixed-input bound is numerically zero. For $S_{331}$ at the same leakage and $S_{331}\simeq5.1961$, the corresponding bounds are approximately $0.087$ bits for the average estimator and $0.003$ bits for the fixed-input estimator. These comparisons are made at witness values close to the strict-qubit maxima, rather than at the maximal almost-qubit values for the specified leakage.

The different requirements for perfect guessing help explain the greater robustness of the average estimator. For fixed-input generation, only the selected output must be deterministic in each strategy known to the adversary; the remaining input pairs may retain unpredictable outcomes while contributing to the observed violation. By contrast, the uniform average guessing probability equals one only if every input pair is perfectly predictable in every strategy of nonzero weight. Increasing the predictability of a single pair is therefore insufficient to make the average guessing probability reach one. Although dimensional leakage can assist the adversary in both cases, the average objective requires predictability across the full input ensemble.

This interpretation does not imply a universal ordering between the two estimators. The average guessing probability is optimized over a common adversarial decomposition across all inputs, rather than obtained by averaging independently optimized fixed-input bounds. The results for $S_3$ and $S_{331}$ show that averaging can improve robustness to dimensional leakage and extend the range of violations compatible with certified randomness, as observed for $S_4$ in the main text. A suitably chosen fixed input can nevertheless remain preferable near the ideal quantum maximum when leakage is sufficiently small.

\section{A tight classical bound for $S_4$}
\label{app:classical}
For completeness, we derive the classical bound of $S_4$ under the average leakage assumption. For each $\lambda$, we define the leakage of preparation $x$ as
\begin{equation}
    \ell_x^\lambda
    =1-\sum_{m\in\mathcal M_2^\lambda}p(m|x,\lambda)
    =\sum_{m\notin\mathcal M_2^\lambda}p(m|x,\lambda),
\end{equation}
so that Eq.~\eqref{eq:classical_leakage} is equivalent to $\sum_\lambda q(\lambda)\ell_x^\lambda\leq\varepsilon$ for every $x$. 

For a fixed $\lambda$, we split the contribution to $S_4$ according to whether the communicated message lies inside or outside the alphabet $\mathcal M_2^\lambda$. The part of the behavior that stays inside $\mathcal M_2^\lambda$ cannot exceed the usual bit bound $S_4=4$. On the other hand, whenever a preparation leaks outside this alphabet, its contribution to $S_4$ can improve by at most $2$, since each preparation appears in two correlators, whose total algebraic value is at most $2$. Therefore, for each $\lambda$,
\begin{equation}
    S_4^\lambda\leq4+2\sum_x\ell_x^\lambda,
\end{equation}
and averaging over the shared randomness gives
\begin{equation}
    S_4
    \leq4+2\sum_x\sum_\lambda q(\lambda)\ell_x^\lambda
    \leq4+8\varepsilon,
\end{equation}
where in the last step we used the same leakage bound $\varepsilon$ for the four preparations. Combining this with the algebraic maximum $S_4\leq8$, we obtain
\begin{equation}
    S_4^{\mathrm C}(\varepsilon)
    \leq\min\{4+8\varepsilon,8\}.
\end{equation}

To complete the proof, we show that the bound is tight (i.e. can be achieved). For $\varepsilon=0$, the usual bit strategy gives $S_4=4$. At $\varepsilon=1/2$, shared randomness can be used so that, for each value of $\lambda$, two preparations belong to the two-symbol alphabet and the other two leak, while averaging over $\lambda$ makes every preparation leak with probability $1/2$. In this case the algebraic value $S_4=8$ can be reached. By convexly mixing this strategy with the usual bit strategy, one obtains all intermediate values $S_4=4+8\varepsilon$ for $0\leq\varepsilon\leq1/2$. Therefore, $S_4^{\mathrm C}(\varepsilon)=\min\{4+8\varepsilon,8\}$.

\section{Fock-space leakage of a Gaussian state}
\label{app:gaussian_fock}
For two single-mode Gaussian states $\rho_0$ and $\rho_1$, with displacement vectors $\boldsymbol d_0,\boldsymbol d_1$ and covariance matrices $V_0,V_1$, the fidelity is given in Sec.~III.A.3, Eq.~(80) of Ref.~\cite{Weedbrook2012} by
\begin{equation}
\begin{aligned}
F(\rho_0,\rho_1)
&=
\frac{2}
{\sqrt{\Delta+\delta}-\sqrt{\delta}}
\\
&\quad\times
\exp\left[
-\frac12
(\boldsymbol d_0-\boldsymbol d_1)^T
(V_0+V_1)^{-1}
(\boldsymbol d_0-\boldsymbol d_1)
\right].
\end{aligned}
\end{equation}
where $\Delta=\det(V_0+V_1)$ and
$\delta=(\det V_0-1)(\det V_1-1)$.

We now take $\rho_0=\rho$ and $\rho_1=|\alpha\rangle\langle\alpha|$, with $V = V_0$ and $\boldsymbol d=\boldsymbol d_0$. In the convention $[\hat q,\hat p]=2i$, a coherent state has covariance matrix $V_1=\mathbbm{1}$ and displacement $\boldsymbol d_1 =\boldsymbol d_\alpha=(2\operatorname{Re}\alpha,2\operatorname{Im}\alpha)^T$ (see Sec.~II.B.2 of Ref.~\cite{Weedbrook2012}). Since $\det V_1=1$, we have $\delta=0$, while $\Delta=\det(V+\mathbbm{1})$. Moreover, since the coherent state is pure,
$F(\rho,|\alpha\rangle\langle\alpha|)=\langle\alpha|\rho|\alpha\rangle$.
Defining $A=V+\mathbbm{1}$ and $J=A^{-1}$ therefore gives
\begin{equation}
    \langle\alpha|\rho|\alpha\rangle
    =
    \frac{2}{\sqrt{\det A}}
    \exp\left[
    -\frac12
    (\boldsymbol d_\alpha-\boldsymbol d)^T
    J
    (\boldsymbol d_\alpha-\boldsymbol d)
    \right].
    \label{eq:gaussian_coherent_overlap}
\end{equation}

Using $|\alpha\rangle=e^{-|\alpha|^2/2}\sum_{n=0}^{\infty}\alpha^n|n\rangle/\sqrt{n!}$, we have
\begin{equation}
    e^{|\alpha|^2}\langle\alpha|\rho|\alpha\rangle
    =
    \sum_{m,n=0}^{\infty}
    \frac{(\alpha^*)^m\alpha^n}{\sqrt{m!n!}}
    \langle m|\rho|n\rangle,
\end{equation}
from which the first two Fock populations follow as
\begin{align}
    p_0
    &=
    \langle0|\rho|0\rangle
    =
    \left.
    \langle\alpha|\rho|\alpha\rangle
    \right|_{\alpha=0},
    \\
    p_1
    &=
    \langle1|\rho|1\rangle
    =
    \left.
    \partial_{\alpha}\partial_{\alpha^*}
    \left[
        e^{|\alpha|^2}
        \langle\alpha|\rho|\alpha\rangle
    \right]
    \right|_{\alpha=0},
\end{align}
since $\partial_\alpha\partial_{\alpha^*}$ extracts the coefficient of $\alpha\alpha^*$, while setting $\alpha=0$ removes all terms containing higher powers. The first expression gives
\begin{equation}
    p_0
    =
    \frac{2\exp[-\boldsymbol d^T J\boldsymbol d/2]}
    {\sqrt{\det A}}.
\end{equation}
For the second, write $\alpha=u+iv$ and $\boldsymbol s=(u,v)^T$, such that $\boldsymbol d_\alpha=2\boldsymbol s$, $|\alpha|^2=\boldsymbol s^T\boldsymbol s$, and $\partial_\alpha\partial_{\alpha^*}=(\partial_u^2+\partial_v^2)/4$. Using Eq.~\eqref{eq:gaussian_coherent_overlap}, we can write
\begin{equation}
    e^{|\alpha|^2}\langle\alpha|\rho|\alpha\rangle
    =
    \frac{2}{\sqrt{\det A}}\,e^{g(\boldsymbol s)},
\end{equation}
with
\begin{equation}
    g(\boldsymbol s)
    =
    \boldsymbol s^T(\mathbbm{1}-2J)\boldsymbol s
    +2\boldsymbol s^T J\boldsymbol d
    -\frac12\boldsymbol d^T J\boldsymbol d .
\end{equation}
At $\boldsymbol s=0$ ($\alpha =0$),
\begin{equation}
    \nabla g
    =
    \begin{pmatrix}
        \partial_u g\\
        \partial_v g
    \end{pmatrix}
    =
    2J\boldsymbol d,
    \quad
    (\partial_u^2+\partial_v^2)g
    =
    4(1-\operatorname{Tr}J).
\end{equation}
Using
$(\partial_u^2+\partial_v^2)e^g
=e^g[(\partial_u^2+\partial_v^2)g+|\nabla g|^2]$,
we then obtain
\begin{equation}
    p_1
    =
    p_0
    \left(
        1-\operatorname{Tr}J
        +\boldsymbol d^T J^2\boldsymbol d
    \right).
\end{equation}
\end{document}